\documentclass[psfig]{./elsart}
\newcommand{\msun}{\mbox{${\rm M}_\odot$}}
\newcommand{\rsun}{\mbox{${\rm R}_\odot$}}

\newcommand{\kms}{\mbox{${\rm km~s}^{-1}$}}
\newcommand{\SeBa}{\mbox{\sf SeBa}}
\newcommand{\A}{\mbox{\em A}}
\newcommand{\B}{\mbox{\em B}}

\def\TZ{Thorne-$\dot {\rm Z}$ytkow}
\def\TZO{T$\dot {\rm Z}$O}
\def\apgt{\ {\raise-.5ex\hbox{$\buildrel>\over\sim$}}\ }
\def\aplt{\ {\raise-.5ex\hbox{$\buildrel<\over\sim$}}\ }

\def\eqlt{\ {\raise-.9ex\hbox{$\buildrel<\over-$}}\ }

\begin{document}
\begin{frontmatter}
%\thesaurus{08.02.1; 08.05.3; 08.14.1; 08.19.4}
\title{The origin of single radio pulsars}
%\thanks[JSPS]{Japan Society for the
%		 Promotion of Science Fellow}
\thanks[HF]{Hubble Fellow}
\author[Tokyo,BU]{Simon~F.~Portegies~Zwart\thanksref{HF}}
\author[Adam,ITP]{Edward~P.~J.~van~den~Heuvel}
%\offprints {S.~F.~Portegies~Zwart: spz@grape.c.u-tokyo.ac.jp}
\address[Tokyo]
	   {Deptartment of General System Studies,
	   University of Tokyo, 3-8-1 Komaba,
	   Meguro-ku, Tokyo 153, Japan}
\address[BU]
	   {Deptartment of Astronomy
	    Boston University,
	    725 Commonwealth Avenue,
	    Boston, MA 01581, USA}
\address[Adam]{Astronomical Institute {\em Anton Pannekoek}, 
           Kruislaan 403, 1098 SJ Amsterdam, The Netherlands}
\address[ITP]
 	   {Institute for Theoretical Physics UCSB, Santa Barbara, 
	   California 93106-4030, USA}

%\date{Received; accepted: }
%%\authorrunninghead{S.\ F.\ Portegies Zwart \& E.~P.~J.~van~den~Heuvel}
%\maketitle
%\markboth{S.~F.~Portegies~Zwart \& Edward~P.~J.~van~den~Heuvel}
%	 {The origin of single radio pulsars}

\begin{abstract}
By comparing relative rates of supernovae versus formation rates of
single radio pulsars, recycled pulsars, binary pulsars and X-ray
binaries we put strong limits on the progenitors of radio pulsars and
on the requirement of an asymmetry in the supernova.  The assumption
that radio pulsars are only formed in type Ib and type Ic supernovae
from interacting binaries (Iben \& Tutukov 1996) breaks down on the
implication that in that case either the formation rate of binary
pulsars (double neutron stars) should be of the order of 20\%\, of the
single pulsar birthrate or, alternatively, almost all single pulsars
(85\% to 98\%) should originate from \TZ\ stars. In the latter case
the pulsar velocity distribution is inconsistent with observations.
Also, in that case the difference between the supernova rate and the
pulsar formation rate would be about one order of magnitude, i.e.:
much larger than observed.

Allowing type~II supernovae from single stars and non-interacting
binaries to form radio pulsars solves this conundrum, but then a kick
is required in order to explain the high velocities of single radio
pulsars.  A kick is also required to understand the small birthrate,
relative to the supernova rate, of binary pulsars consisting of two
neutron stars.

%\keywords{stars:      binaries    -- 
%	  stars:      evolution   -- 
%	  stars:      neutron     -- 
%	  stars:      supernovae}    
\end{abstract}
\end{frontmatter}

\section{Introduction}
Neutron stars are believed to descend from stars which are massive
enough to experience a supernova at the end of their fuel processing
lifetime.  However, it is not completely clear whether or not all
massive stars finally produce a radio pulsar (i.e.: a highly
magnetized and rapidly
rotating neutron star); in some cases the
remnant may not show up as a radio pulsar and above certain mass
limits the star may collapse to a black hole (van den Heuvel \&
Habets 1984; Portegies~Zwart et al.~1997a; Ergma \& van den Heuvel
1998) or the star may completely detonate and leave no remnant at all
by a pair creation explosion.  The physical parameters at the moment
of the supernova which are required to form a radio pulsar are not yet
very clear, but strong limits can be set on the possible progenitors.

According to the most simplistic picture, we recognize three types of
supernovae: type Ia, Ibc and type II which are of importance for the
argumentation which we set out in this paper (see e.g.\ Nomoto et al.\
1995; Thielemann et al.\ 1996).  A type II supernova is the result of the
collapse of the core of a single star or a component of a wide,
non-interacting binary, with a mass larger than $9 \pm
1$~\msun\ that still has a hydrogen envelope at the time of the
collapse (Timmes et al.\ 1996; Iben et al.\ 1997).  A star in a binary
with an orbital separation so large that it evolves unaffected is
considered single in this respect.

A type Ib or Ic supernova is thought to be generated by a massive star
which, under the 
influence of another star or due to a strong stellar wind (i.e.\
initial mass $\apgt 35$~\msun), has lost its hydrogen envelope
(subclass Ib) or in addition also its helium envelope (subclass Ic).
(Type Ia supernovae are generally assumed to have a 
different origin and do not leave a compact star,
we therefore ignore them here cf.\ Canal et al.\ 1997.) 

Pulsars appear to be high-velocity objects. A careful analysis of the
measured proper motions of pulsars indicates a mean characteristic
velocity at birth of order 250 to 300 km/s, with a possible flat
distribution towards low velocities and a tail extending to
$>~800$~km/s (Lyne \& Lorimer 1994; Hartman 1997, but see also Hansen
\& Phinney 1997; Cordes and Chernoff 1997; Lorimer et al.\ 1997).
These high peculiar velocities of single radio pulsars (some 10\% have
$v > 600$~km/s) suggest that there is a mechanism which gives the
newly born pulsar a push. In todays literature two models for this
push are most favored:
\begin{itemize}

\item[\A] Rapidly rotating young radio pulsars are born only from type
Ib and Ic supernovae in binaries and mass loss in the supernova
unbinds the binary (the so called Blaauw mechanism, Blaauw 1961,
1964); the radio pulsar is ejected with its orbital velocity and the
neutron stars that are formed in this way are the only ones that spin
rapidly enough to be observed as radio pulsars; neutron stars
originating from single stars or wide binaries rotate too slowly to
produce radio pulsars. This model for explaining pulsar velocities was
proposed by Tutukov et al.\ (1984) and worked out in detail by Tutukov
\& Yungelson (1993) and Iben \& Tutukov (1996).

\item[\B] An asymmetry in the supernova results in a ``velocity kick''
imparted to the newly born pulsar (Shklovskii 1970; Gunn \& Ostriker
1970, Dewey \& Cordes 1987).  An 
asymmetry of a few percent suffices to explain the observed peculiar
velocities (Woosley 1987; Woosley \& Weaver 1992).  The origin of the
kick can be an asymmetry in the neutrino out flow from
the newly born radio pulsar (Janka \& M\"uller 1994; Herant et al.\, 1994) or
an off center detonation (Burrows \& Hayes 1996).  
The reasons, however, why such asymmetries occur are not understood from a
theoretical point of view.
\end{itemize}

In this paper we argue, using simple estimates and the results of
detailed population synthesis, that type II supernovae, 
(i.e.: supernov\ae\, from stars that have {\bf not} lost their
hydrogen envelopes, and that may be single or in wide binaries)
are required to
add to the formation of radio pulsars and that intrinsic kicks are
most favored to explain the observed characteristics of the population
of radio pulsars.

A number of arguments for the occurrence of kicks is summarized by van
den Heuvel \& van Paradijs (1997, see however, Iben and Tutukov 1998,
for an alternative view) and a lower limit to the velocity of the kick
is provided by Portegies Zwart et al.\ (1997b).  Kalogera \& Webbink
(1998) show that without kicks it is not possible to produce low-mass
X-ray binaries with an orbital period smaller than a day (see,
however, Iben et al.\ 1995 who report to have no difficulty producing
low-mass X-ray binaries in the absence of kicks). Tauris \& Bailes
(1996) demonstrate that it is difficult to produce millisecond pulsars
with a velocity $\apgt 270$\kms\ without an asymmetric kick.
Asymmetric velocity kicks in supernovae are also favored in various
population synthesis calculations pioneered by Dewey \& Cordes (1987)
who showed that without kicks many more double neutron stars would be
produced than are observed (see also Meurs \& van den Heuvel 1989;
Dalton \& Sarazin 1995; Portegies Zwart \& Spreeuw 1996; Lipunov et
al.\ 1996; Lipunov et al.\ 1997; Terman et al.\ 1998).

\section{A simple analytical consideration}

\subsection{Birthrates without kicks}

If all stars are born single and there are no binaries in the galaxy,
scenario \A\ implies that no radio pulsars are formed at all. In other
words, this scenario for the formation of single radio pulsars
excludes a star which is born single as a progenitor.  The majority of
the observed well-studied stars is member of a binary system anyway,
so this poses no direct problem for scenario \A.  For simplicity we
will now assume that all stars are born in binaries and that model \A\
is correct: supernovae are symmetric (no velocity kick is given to the
stellar remnant which is formed in the supernova) and only stars which
have lost their envelopes in the interaction with a companion star
produce radio pulsars in the supernova, as Iben \& Tutukov
(1996) have proposed.

The requirement for the formation of a single radio pulsar is then
that the binary must 1.) experience a phase of mass transfer or
common-envelope evolution, and 2.) is dissociated in the first or the
second supernova, or alternatively 3.) completely spiral-in, in a
phase of mass-transfer following the first supernova producing a \TZ\
object (Thorne \& Zytkow 1975; 1977) that leaves a single pulsar as a
remnant (Podsiadlovski et al.\ 1995; Iben \& Tutukov 1996). We return to the
uncertainty of forming a \TZ\ star at the end of this \S.

In practice this type of evolution will happen only to a small subset
of all binaries. A binary with a very short orbital period and/or a
small mass ratio will not survive the first phase of mass transfer and
merges into a single object. Such a single star is, according to Iben
\& Tutukov (1996) no candidate for producing a radio pulsar. Neither
are the binaries which are initially too wide to experience a phase of
mass transfer. Only the binaries in the range of orbital periods
between several days and a few decades are consequently candidates for
producing single radio pulsars.

Our population synthesis calculations given in \S\,\ref{Sect:popsynth}
demonstrate that the majority of
binaries that experience and survive their first mass transfer or
first common-envelope phase stay bound after the first (symmetric)
supernova (see \S\,\ref{Sect:popsynth}).  Only those binaries for which the
mass which is lost in the supernova exceeds half the total binary mass
prior to the supernova are dissociated.  With conservative mass
transfer it is always the lowest mass component which explodes first,
and no systems are disrupted.  Even if the initial mass-ratio
distribution strongly favors small mass ratios and if mass is not
conserved in the binary system during mass transfer or common-envelope
evolution the unbound fraction is still small.  A simple way to
see this is to consider the shape of the initial mass function: stars
between $\sim 8$~\msun\ and $\sim 15$~\msun\ contribute about half of all
supernova progenitors; they leave helium cores with a mass smaller
than $\sim 3.8$~\msun\ after the first mass transfer or common-envelope
phase. Since companions $\aplt 1$\,\msun\ will spiral in completely
and coalesce, 
and neutron stars have a mass of about 1.4\,\msun, the systems that
survive the first mass transfer then lose less than half the total
mass and therefore remain bound after the first supernova explosion.
Since the masses of companions of stars
$> 15$~\msun\ are in most cases expected to be considerably more
massive than 1~\msun, also a large fraction of the systems with more
massive primaries remain bound.
Therefore, the majority of the binaries that survive the first phase
of mass transfer will remain bound after the first supernova explosion.
The fraction that is disrupted in the first supernova (of binaries
that experience and survive the first mass transfer), we denote as
$y$. A conservative estimate of $y$ is: $y < 0.25$ (Our simulations in
the next section (\S\,\ref{Sect:popInt}) show $y$ to be $\aplt
0.1$.).

A binary which survives the first supernova explosion 
becomes a high-mass X-ray binary as soon as the companion of the neutron star
starts to transfer mass. 
Most of these systems will go through a Be/X-ray binary phase (see
e.g.:  van den Heuvel and Rappaport 1987). The neutron star is spun up
in this phase and 
it may become a recycled pulsar.

Subsequently a fraction of the 
binaries where a neutron star accretes from its companion will
spiral-in in a common-envelope phase and merge to form a \TZ\ Object.
Of the systems that survive as binaries after the spiral-in, only those
will be disrupted in a symmetric explosion for which the exploding
helium star is more massive 
than 4.2~\msun.  
We will denote the fraction of systems that survive the second
supernova as binaries as $\alpha$.
The dissociated binary ejects two radio pulsars; one
young and one recycled.  Again with the initial-mass function
argument, of the order of half of these helium-star binaries have
companions to the neutron stars that have helium cores $\aplt
4.2$~\msun, and therefore about half of these systems will 
not be disrupted in the second (symmetric) supernova\footnote[1]{
The phase of mass transfer which preceded the first supernova
affects the secondary mass and the initial-mass function argument
cannot be applied trivially; the
mass transfer process has increased the secondaries mass and therefore
the mass of its core. However, taking this into account, still not
more than half the systems are expected to be disrupted in the
symmetric second super nova.}.

According to Iben \& Tutukov's (1996) model there are then three types
of radio pulsars originating from high-mass X-ray binaries: 1.)
single pulsars resulting from binaries disrupted at the second
supernova [producing two pulsars]; 2.) single pulsars resulting from
complete spiral-in of high-mass X-ray binaries, to form a \TZ\ Object
and then a recycled pulsar; 3.) double neutron stars.  We chose $y$
to be the fraction of post mass-transfer systems which are
disrupted in the first supernova explosion.  
So, if a fraction $x$ of all high-mass X-ray binaries
spiral in completely to form \TZ\ objects and then single pulsars,
the fraction $(1-x)$ of high-mass X-ray binaries that survive the
spiral in will leave helium
star plus neutron star binaries, producing $\alpha (1-x)$ double
neutron stars and $ 2 (1-\alpha) (1-x)$ single pulsars.
As the X-ray binaries formed a fraction $(1-y)$ of all post
mass-transfer systems one thus will have that the
fraction of double neutron stars among all pulsars is
\begin{equation}
%\frac{0.5(1-x)(1-y)}{1+0.5(1-x)(1-y)}. \nonumber
\frac{\alpha (1-x)(1-y)}
     {y + x(1-y) + 2(1-\alpha)(1-x)(1-y)}. 
\label{Eq:fraction}\end{equation}
The observed fraction of double neutron star among the entire pulsar
population is about $0.6$\% ($\sim 6$ binary pulsars among $\sim 1000$
single pulsars).  Assuming $y = 0.25$ we then obtain, for
$\alpha=0.5$, that $x=0.984$.
Pulsars in close binaries are probably under represented because they
are plagued by extra selection effects due to the acceleration of the
pulsar in the binary (Johnston \& Kulkarni 1991).  
If the
real fraction of double pulsars would be an order of magnitude larger
than observed (i.e.: 6 per cent) then, with $\alpha=0.5$, 
 still we obtain $x=0.849$,
i.e.: more than 85\% of all pulsars would decend from \TZ\, objects. 

Scenario \A\ would therefore
imply that between  85 and 98 per cent of all radio pulsars descend
from \TZ\ objects, which is an absurd result.
(Even in the very unrealistic case that in a symmetric explosion only
20 per cent of the helium-star plus neutron star binaries would
survive the second supernova explosion, still an ``observed'' 6 per
cent of binary pulsars would imply that more than half of all radio
pulsars have decended from \TZ\, objects.)

Moreover as the bulk of the high-mass X-ray binaries which produced
the \TZ\, objects are Be-type X-ray
binaries, which have small runaway velocities ($11\pm 6.7$~km/s;
Chevalier \& Ilovaisky 1998), between 85 and 98 per cent of the pulsars
would, according to the model of Iben \& Tutukov (1996) be very 
low-velocity objects, contrary to the observations.

There might possibly be an alternative evolutionary path in the case
of symmetric supernovae to avoid these contradictions as follows: the
suggestion provided by Chevalier (1993; see also Bisnovatyj-Kogan \&
Lamzin 1984; Fryer et al.\ 1996; Brown \& Bildsten 1998) that a
neutron star in a common envelope may accrete hyper critically and
transforms to a black hole.  In this case the old neutron star does
not become a recycled pulsar but collapses into a black hole instead.
In that case, if the binary survives the common-envelope phase
altogether, a high kick velocity is required to dissociate the binary
upon the second supernova; the higher mass of the black hole easily
prevents dissociation of the binary in a symmetric supernova.
Scenario \A\ (with no kicks) thus predicts in this case that, while
the birthrate of double neutron stars is small, many young pulsars
should be accompanied by a black hole in a short period orbit, which
is obviously contradicted by the observations.  Furthermore in this
case, also \TZ\ objects will always produce black holes, so this
channel for pulsar formation is lost.

Thus already from these 
simple analytical considerations one observes that
with symmetric supernova explosions either many binary radio pulsars
with black holes are
produced or between 85 and  98 per cent of
all pulsars must result from \TZ\ objects and will have low space
velocities -- in complete disagreement with the observations.

We will now show that population-synthesis calculations completely
confirm the results from the analytical calculations.

\section{Results from population synthesis}\label{Sect:popsynth}

For the 
numerical simulations we use the binary evolution program \SeBa\
(see Portegies Zwart \& Verbunt 1996) with more than a million
binaries with a primary mass between 8~\msun\ and 100~\msun\ selected
from a power-law distribution with exponent 2.5 (Salpeter $=
2.35$). All binaries are evolved in time until the second supernova
occurs (see Portegies Zwart \& Yungelson 1998 for a detailed
description of the models and initial conditions).  We assume that all
stars are born in binaries with a semi-major axis up to
$a=10^6$~\rsun\ to be present in a flat distribution in $\log a$
(Duquennoy \& Mayor 1991). The mass of the secondary is selected
between 0.1~\msun\ and the mass of the primary from a distribution
flat in mass ratio (Hogeveen 1992).  The results of these computations are
summarized in Tab.~\ref{Tab:psrform}.  The results are presented in
three decimals in order to make the numbers recognizable.  In practice
the last decimal may easily be omitted due to the uncertainties in
initial conditions, physics and model parameters.  We consider cases
with and without kicks and now discuss the outcome for the different
models for pulsar formation mentioned above.

\begin{table}
\caption[]{ The number of neutron stars formed, normalized to the total
number of supernovae for two models; without a kick (columns two and
three; see model $A$ from Portegies Zwart \& Yungelson 1998) and in
which a kick is imparted to the newly formed neutron star (columns
four and five; see their model $B$).  The first column identifies the
system which results from the supernova. Notation is taken from
Portegies Zwart \& Verbunt (1996): {\em ns}\ stands for a neutron star
and $\star$ for any non-remnant star, parenthesis `(~,~)' indicate a
detached but bound binary and braces `\{~~\}' a merged object. \\
The second column gives the relative fraction of the various systems
which originate from exploding naked Helium or Carbon-Oxygen stars
(supernovae type Ib and Ic) the third column gives the results for
supernova type II (single stars or stars in wide binaries which have
lost their hydrogen envelopes by 
their own radiation pressure in a Wolf-Rayet phase are also included
in this column and denoted as type II supernovae as according to Iben
\& Tutukov, 1996, these do not contribute to pulsar formation).  The
results for 
the model with a kick (according to the distribution from Hartman
1997) are presented in the last two columns.  The data for the
binaries which are dissociated upon the second supernova include also
the binaries where the primary produced a black hole. This pollution,
however, is not that big; $\sim 5$\% and $\sim 22$\% for the models
without and with a kick, respectively. \\
The total does not add to unity because some supernovae produce
black holes instead of neutron stars; note that in the model without a
kick $\sim 20$\% more black holes are formed than in model $B$.  The
last row presents the formation rate of \TZ\ objects relative to the
total supernova rate.  
}
\begin{tabular}{l|lr|lr}
\noalign{\smallskip}
\hline
\noalign{\smallskip}
&\multicolumn{2}{c}{Without kick}
&\multicolumn{2}{c}{With kick} \\
\noalign{\smallskip} 
result & SN Ibc &  SN II & SN Ibc & SN II  \\ 
       &[naked]& [+WR]  &[naked]& [+WR]  \\ \hline
\noalign{\smallskip} 
&\multicolumn{4}{c}{After the first supernova} \\
\{{\em ns}\}		    & 0.003 & 0.157 & 0.003 & 0.155\\
({\em ns}, $\star$)      & 0.075 & 0.244 & 0.025 & 0.012 \\
{\em ns}, $\star$        & 0.012 & 0.241 & 0.061 & 0.469 \\ 
&\multicolumn{4}{c}{After the second supernova} \\
{\em ns}		    & 0.001 & 0.055 & 0.004 & 0.211 \\ 
({\em ns}, {\em ns})          & 0.011 & 0.026 & 0.002 & 0.000 \\
{\em ns}, {\em ns}            & 0.017 & 0.104 & 0.004 & 0.008 \\ %%)ns, ns( +
						       %%)bh, ns(
\hline
\noalign{\smallskip} 
Total:              & 0.119 & 0.827 & 0.100 & 0.855 \\
\TZO\   &\multicolumn{2}{c}{0.001} & \multicolumn{2}{c}{0.004} \\ \hline 
\noalign{\smallskip}
\hline
\end{tabular}
\label{Tab:psrform}\end{table}

\subsection{Single and double pulsar formation rate if only type Ib 
and Ic supernovae produce pulsars} \label{section_a}

\subsubsection{The case of symmetric mass ejection}\label{Sect:popNK}
 
In the model which does not
incorporate a velocity kick the fraction of type Ibc supernovae from
binaries which produce a neutron star to the total number of type Ibc
+ II supernovae is $\sim 11.9\%$ (see Tab.~\ref{Tab:psrform}).

A binary which survives the first phase of mass transfer becomes a
({\em he}, $\star$) binary (see the table caption for an explanation of the
notation).  If such a binary is disrupted in the first (type Ibc)
supernova (which occurs in 1.2\% of all supernovae) a single {\em ns}\
(pulsar) and a single $\star$ are released.  In some rare cases the
binary experiences, and survives, two phases of mass transfer before
the first supernova occurs and becomes a double helium star ({\em he},
{\em he}) binary.  Dissociation of such a ({\em he}, {\em he})\ binary
upon the first supernova releases, next to a single pulsar, a single
helium or Carbon-Oxygen star which may explode at a later
instant. Since this single helium star has lost its hydrogen envelope
due to the interaction with its companion it is, according to Iben \&
Tutukov (1996), also a candidate for the formation of a single radio
pulsar contributing with a modest 0.1\%.  A ({\em he}, {\em he})
binary which experiences an additional phase of mass transfer before
the first supernova occurs may merge and become a single rapidly
rotating helium or Carbon-Oxygen star.  The explosion of this single
helium star in a type Ibc supernova contributes with 0.3\% to the
pulsar formation rate as fraction of the total supernova rate (see
Tab.~\ref{Tab:psrform} after \{{\em ns}\}).

A ({\em he}, $\star$) binary which remains bound after the first type Ibc
supernova but is dissociated upon the second supernova releases two
pulsars and contributes with $2\times 1.7\% = 3.4\%$ to the production
rate of single neutron stars: both are pulsars.  \TZ\ objects
contribute only little ($\sim 0.1$~\%) to the pulsar formation rate
(see Tab.~\ref{Tab:psrform}).  Assuming only type Ib supernovae to
produce pulsars, the total number of single radio
pulsars produced as a fraction of the total number of supernovae (type
Ib, Ic and type II together) according to this model is $\sim 5.1\%$ ($\equiv 1.2 +
0.1 + 0.3 + 2\times 1.7 + 0.1$).  
As to the double pulsar (neutron star binary) formation rate,
Tab.\ref{Tab:psrform} shows that in the model without kicks one
expects 0.011 double ones relative to 0.051 single ones, hence about
20 percent of all pulsars is expected to be born double.

\subsubsection{Comparison with other population synthesis results}
For our models without kicks the fractions of pulsars produced from
type Ibc supernovae relative to the total supernova rate is much
smaller than that derived by Iben and Tutukov (1996) who find a
birthrate for radio pulsars of 0.007 per year relative to a total
birthrate for neutron stars of 0.028 per year; i.e.: 25\% of all
supernovae produce a single radio pulsar (see also Tutukov \&
Yungelson 1993).  At least part of this discrepancy is a result of the
difference in the fraction of binaries which experience mass transfer
during their lifetime.  In the calculations of Iben and Tutukov a
relatively large fraction of binaries experience mass transfer at some
time during their evolution and the contribution of type~II to the
total supernova rate is therefore considerably smaller.  We can
estimate this effect from the results in Tab.~\ref{Tab:psrform} by
counting only the binaries which experience a phase of mass transfer
and re-normalizing our results to the type Ibc supernova rate.

%The lack of a description of the model of Iben \& Tutukov (see however
%Tutukov \& Yungelson, 1993, for a more detailed description) makes it
%hard to understand the origin of this difference, but at least part of
%it is a result of the difference in the assumed maximum to the initial
%separation for a zero-age binary: Iben \& Tutukov chose 1000~\rsun\ as
%a maximum where we use $10^6$~\rsun.  Since most of the stars in
%binaries with a separation up to 1000~\rsun\ interact during evolution
%the type Ibc supernova rate is therefore in their relative model
%without kicks much larger than ours, and their type~II supernova rate
%is considerably smaller.

\subsubsection{Formation rates from interacting binaries}\label{Sect:popInt}

In a population where all binaries transfer mass at some stage during
their evolution the only source for type II supernovae is formed by
binaries which merge before the first supernova and explode as single
stars (0.157 for our model without a kick), and from ({\em he}, $\star$) binaries
which are dissociated upon the first type Ibc supernova explosion
(1.2\%), i.e.: of which the secondary may explode as if the star was
born single. 
We computed in section\,\ref{Sect:popNK} in case of no kicks, the
fraction of type Ibc supernova to the total supernova rate is 11.9\%.
The contribution of type Ibc and type II supernovae from interacting
binaries to the total supernova rate (including the non-interacting
binaries) is in our model therefore given by $\sim 0.288$ ($\equiv
0.119 + 0.157 + f \times 0.012$); i.e.: $\sim 29$\% of all supernovae
originate from interacting binaries.  The fraction $f$ ($\simeq 0.92$)
is introduced to quantize the fraction of ({\em he}, $\star$) binaries which is
dissociated upon the first supernova and of which the released
companion may explode in a type II supernova).  The formation rate of
single pulsars formed in type Ibc supernovae as a fraction of the
supernova rate in interacting binaries then becomes $5.1\%/0.29 \approx
18$\%. This rate is of similar order as the result of Iben and Tutukov
(1996) who derive a fraction of 25\%.  This may indicate they they
underestimate the contribution of wide binaries to the supernova
rate. Note, however, that we underestimated the contribution to type
II supernovae due to our adopted minimum mass of 8\,\msun\ to the
initial primary mass (Iben \& Tutukov adopted a minimum of 10\,\msun).
A binary, for example, which contains a 7\,\msun\ and a 4\,\msun\ star
that merges in the first, unstable, phase of mass transfer might form
a single star which is massive enough to explode; these binaries are
not accounted for in our simulation.  By comparing our results with
those of Portegies Zwart \& Verbunt (1996, see their Tab.~4), who also
take lower mass binaries into account, we estimate that this effect
contributes with $\aplt 10$\% to the total supernova rate.

\subsubsection{Birthrates with kicks}

Following the same analysis for the model in which a velocity kick is
imparted to the newly born neutron star, the total number of single
pulsars produced if only type Ib,c supernovae produce pulsars is 8\%
($\equiv 0.3 +6.1 +0.4 +2\times0.4$ + 0.4) and the pulsar formation rate from
interacting binaries among the total super nova rate becomes between
25\% and 31\% [ $8\%/(0.100 +0.155+ f \times 0.061)$] (see table
1). In contrast to the models without a kick, a considerable fraction
of the ({\em he}, $\star$) binaries is dissociated by the first type Ibc supernova
(71\%) and as a consequence the contribution of the released secondary
stars to the type II supernova rate is considerable. In the model with
kicks in which only Type Ibc supernovae produce pulsars, the fraction
of binary pulsars produced is 0.002/0.08, i.e. about 2.5\%, i.e.\ some 8
times lower than in the case without kicks.

\subsection{Discrepancy between pulsar formation rate and supernova rate
in case pulsars originate only from Type Ibc supernovae}

The model without a kick in which only type Ibc supernovae produce
pulsars predicts a discrepancy between the observed supernova rate (of the
order of $\sim 0.012$ type II per year and $\sim 0.002$ type Ibc per
year, see Cappellaro et al.\ 1997) and the 
single-pulsar formation rate (only 5.1\% of the total supernova rate,
see \S\,\ref{section_a} above) of a factor 20. This is clearly
contradicted by the observations, which indicate a pulsar formation
rate of the same order as the supernova rate in the Galaxy: 
0.004 to 0.008 per year was derived by Lorimer et al.\ (1993)
and Hartman et al.\ (1997) arrive at a pulsar birthrate of 
$\sim 0.003$ per year in the Galaxy, i.e.: differing by a factor 3 or
less from the supernova rate. 

The existence of wide binaries is confirmed by the observations, and
we use the total supernova rate for interacting as well as the
non-interacting binaries in the further discussion.

In the population synthesis models where also type~II supernovae
produce radio pulsars the discrepancy between the pulsar formation
rate and the supernova rate completely vanishes.  In addition to
the formation rate of single pulsars from type Ibc supernova (0.051
and 0.080 of the total supernova rate for the models without and with
a kick, respectively) type II supernovae make a large contribution to
the single pulsar formation rate as can be seen from the table.
Binaries that merge before the first type II supernova contribute with
15.7\% (15.5\% for the model with a kick) to the formation of single
pulsars.  In non-kick models, non-interacting binaries contribute with
24.1\% upon the first supernova (which dissociates the binary) and
with $\sim 4.3$\% ($\equiv 0.055- f \times 0.012$) from the released
companion which might also experience a supernova (note that the
correction factor $f \times 0.012$ for binaries which are dissociated
upon the first type Ibc supernova and of which the secondary
experiences a type II supernova must be applied again.)  For the model
with a kick these fractions are 46.9\% and $\sim 15.0$\% ($\equiv 0.211-f
\times0.061$) for the first and second type II supernova,
respectively.  A smaller fraction of the binaries are dissociated upon
the second collapse, releasing two pulsars; 10.4\% for symmetric and
0.8\% for asymmetric supernovae. The total contribution of type II
supernovae to the single pulsar formation rate becomes $\sim 65$\%
($\equiv 0.157 + 0.241 + 0.043 + 2 \times 0.104$) for the model without a
kick and $\sim 79$\% ($\equiv 0.155 + 0.469 + 0.150 + 2\times 0.008$) if a
kick is imparted to a newly formed neutron star.
 
If type II supernovae contribute to the formation of single radio
pulsars the models without a kick predict that $\sim 70$\% ($\equiv
0.051 + 0.65$) of all type Ibc plus type II supernovae produce a radio
pulsar, and 3.7\% form double neutron stars (binary pulsars).  Using a
supernova rate of 0.01 per year the birthrate of single radio pulsar
then becomes $\sim 0.007$~per year and the birthrate of binary pulsar
is $\sim 4 \times 10^{-4}$ per year.  For the model with a kick these
fractions are $\sim 87$\% and 0.2\%, respectively, resulting in a
birthrate of single pulsars of $\sim 0.009$~per year and for double
neutron stars of $\sim 2 \times 10^{-5}$ per year.

The observed fraction of binary pulsars is about 0.6\% (6 out of some
1000 pulsars). This fraction can, however, not be simply compared with
the above predicted fractions of binary pulsars, since the latter ones
are young (newborn) pulsars, whereas at least 4 out of the known
double neutron star systems in the galactic disk are recycled
ones, i.e.: which live much longer than new born pulsars, as they spin
down much more slowly. 
%(Also, one of the remaining binary pulsars, PSR\,2303+46, has a massive
%white dwarf companion [van Kerkwijk \& Kulkarni 1999], and should be
%omitted [see Portegies Zwart \& Yungelson, 1999, for how to form such
%a system].) 
With a 0.2\% predicted birthrate of double
pulsars among newborn 
pulsars, one indeed would expect only a few non-recycled double
neutron stars among the 1000 pulsars in the galactic disk. It thus
seems that the observed (very low) fraction of non-recycled double
pulsars is in accordance with the predictions from models with
kicks in which also the type II supernovae produce pulsars.  On the
other hand, the observed fraction of non-recycled double pulsars in
the galactic disk (at least one out of 1000) is some 100 times lower
than that predicted by the model without kicks in which only type Ibc
supernovae produce pulsars, and 37 times lower than predicted by the
model without kicks in which both type Ibc and type II supernovae
produce pulsars.

\section{Conclusions}
From order-of-magnitude estimates 
as well as detailed population synthesis studies we argue
that type II supernovae (i.e.: supernovae resulting from single stars
and components of wide binaries) must contribute to the formation of radio
pulsars in order to explain the similar Galactic rates (within a
factor of a few) of supernovae and birth of single radio pulsars. 
If type II supernovae are excluded from the formation of pulsars, the
predicted pulsar birthrate is at least an order of magnitude smaller
than observed. 

An asymmetry in the supernova is required to satisfy that the
birthrate of high-mass binary pulsars (double neutron stars) is
smaller than the birthrate of single radio pulsar by at least two
orders of magnitude (Bailes 1996).  With symmetric supernovae and pulsars
forming only from interacting binaries one predicts the birthrate of
double neutron stars to be of the order of 20 per cent of the pulsar
birthrate, unless the bulk of the single pulsars ($>85\%$) would
have formed from \TZ\ objects. In the latter case most pulsars should have
low space velocities -- contrary to the observations.

With symmetric supernovae and allowing also pulsar formation from type
II supernovae one still predicts a birthrate of double neutron stars
of about 3.7\% of the supernova rate, four times larger
than observed.  We therefore firmly conclude that, contrary to the
suggestions of Iben \& Tutukov (1996):
\begin{itemize}
\item[--] Single stars and components of wide --non-interacting--
binaries must contribute considerably to the formation of pulsars.
\item[--] Supernova mass ejection is asymmetric, giving a considerable
kick velocity to the neutron star.
\end{itemize}

\bigskip
\bigskip
%\acknowledgements 
{\bf acknowledgments} 
We thank Dipankar Bhattacharya, Lev Yungelson, Icko
Iben and the anonymous referee for critically reading the
manuscript and valuable comments.  We thank the Institute for
Theoretical Physics of the University of California, Santa Barbara,
for its hospitality.  
This work was supported by NWO Spinoza grant
08-0 to E.~P.~J.~van den Heuvel, the JSPS grand to SPZ,
and by NASA through Hubble Fellowship
grant awarded by the Space Telescope Science Institute, which
is operated by the Association of Universities for Research in
Astronomy, Inc., for NASA under contract NAS\, 5-26555.


\begin{thebibliography}{9}
\bibitem{} Bailes, M. 1996, in Compact stars in binaries, ed. J. van
	   Paradijs, E. P. J. van den Heuvel, E. Kuulkers (Dordrecht:
	   Kluwer), p. 213
\bibitem{} Bisnovatyi-Kogan, G.\ S.,  Lamzin, S.\ A.\ 1984, AZh 61, 323
\bibitem{} Blaauw, A.\ 1961, BAIN 15, 265
\bibitem{} Blaauw, A.\ 1964, ARA\&A Vol.\ 2, p.\ 213
\bibitem{} Brown, E.\ F., Bildsten, L.\ 1998, ApJ 496, 915
\bibitem{} Burrows, A., Hayes, J.\ 1996, Phys.\ Rev.\ Lett.\ 76, 325
%\bibitem{} Kaspi, V.\ M.\ 1998, Advances in Space Research, v. 21,
Issue 1, p. 167
\bibitem{} Canal, R.\ Isern, J.\ Ruiz-Lapuente, P.\ 1997, ApJL 488, L35
\bibitem{} Cappellaro, E. Turatto, M., Tsvetkov, D. Yu.\ 1997, 
A\&A 322, 431  
\bibitem{} Chevalier, R.\ 1993, ApJ 411, L33
\bibitem{} Chevalier, R., Ilovaisky, S.\ A.\ 1998 A\&A 330, 201
\bibitem{} Cordes, J. M., Chernoff, D. F.\ 1997, ApJ 482, 971
\bibitem{} Dalton, W.\ W., Sarazin, C.\ L.\ 1995, ApJ 440, 280
\bibitem{} Dewey, R., Cordes, J. M.\ 1987, ApJ 321, 780
\bibitem{} Duquennoy A., Mayor M.\ 1991, A\&A 248, 485
\bibitem{} Ergma, E.\ \& van den Heuvel, E.\ P.\ J.\ 1998, A\&A 
331, L29
\bibitem{} Fryer, C. L., Benz, W., Herant, M., 1996, ApJ, 460, 801
\bibitem{} Gunn, J., Ostriker, J.\ 1970, ApJ 160, 979
\bibitem{} Hansen, B. M. S., Phinney, E. S.\ 1997, MNRAS 291, 569
\bibitem{} Hartman, J. W., Bhattacharya, D., 
	   Wijers, R. A. M. J., Verbunt, F.\ 1997, A\&A 322, 477
\bibitem{} Hartman, J. W.\ 1997, A\&A 322, 127
\bibitem{} Herant, M., Benz, W., Hix, W.\ R., Fryer, C.\ L., Colgate,
S.\ A. 1994, ApJ 435, 339
\bibitem{} Hogeveen S.\ J.\ 1992, Ap\&SS 196, 299
\bibitem{} Iben, I., Jr., Tutu
%\bibitem{} Iben, I., Jr., Tutukov, A. V., Yungelson, L.\ R.\ 1995, ApJS 100, 233
\bibitem{} Iben, I., Jr., Ritossa, C, Garcia-Berro C.\ 1997, ApJ 489, 772
\bibitem{} Iben, I., Jr., Tutukov, A. V.\ 1998, ApJ 501, 263
\bibitem{} Janka, H. -T., M\"uller, E.\ 1994, A\&A 290, 496
\bibitem{} Johnsoton H.\ M., Kulkarni S.\ R. 1991, ApJ 368, 504 
\bibitem{} Kalogera, V.\ \& Webbink, R.\ F.\ 1998, ApJ 493, 351
\bibitem{} Lipunov, V.\ M., Postnov, K.\ A., Prokhorov, M.\ E.\ 
1996, A\&A 310, 489
\bibitem{} Lipunov, V.\ M., Postnov, K.\ A., Prokhorov, M.\ E.\ 1997,
MNRAS 288, 245 
\bibitem{} Lorimer, D.\ R., Bailes, M., Dewey, R.\ J., Harrison, P.\
A.\ 1993, MNRAS 263, 403
\bibitem{} Lorimer, D.\ R., Bailes, M.\ \& Harrison, P.\ A.\ 1997,
MNRAS 289, 592
\bibitem{} Lyne, A.~G.,  Lorimer, D.~R.\ 1994, Nat 369, 127 
\bibitem{} Meurs, E.\ J.\ A.,  van den Heuvel, E.\ P.\ J.\ 1989, A\&A
226, 88
%\bibitem{} Narayan R., Ostriker J.\ 1990, 352, 222
\bibitem{} Nomoto, K., Iwamoto, K., Tomoharu, S.\ 1995, Phys.\ rep.\ 
256, 183 
\bibitem{} Podsiadlowski, P., Cannon, R.\ C., Rees, M.\ J.\ 1995,
MNRAS 274, 485
\bibitem{} Portegies~Zwart, S.~F., Verbunt, F.\ 1996, A\&A 309, 179
\bibitem{} Portegies~Zwart, S.~F., Spreeuw, F.\ 1996, A\&A 312, 670
\bibitem{} Portegies~Zwart, S.~F., Verbunt, F., Ergma, E.\ 1997a,
A\&A 321, 207 
\bibitem{} Portegies~Zwart, S.~F., Kouwenhoven, M., Reynolds, A.\
1997b, A\&A 328, L2   
\bibitem{} Portegies~Zwart, S.~F., Yungelson, L.\ 1998, A\&A 332, 173
\bibitem{} Portegies~Zwart, S.~F., Yungelson, L.\ 1999, MNRAS {\em in press}
\bibitem{} Shklovskii, I.\ S.\ 1970, Astr.\ Zh.\ 46, 715
\bibitem{} Tauris, T.\ M., Bailes, M.\ 1996, A\&A 315, 432
\bibitem{} Terman A.\ L., Taam, R.\ E., Savage, C.\ O.\ 1998, MNRAS
293, 113
\bibitem{} Thielemann, F. K., Hoeflich, P., Khokhlov, A., Nomoto, K.,
	   Wheeler, J. C.\ 1996, AAS 189, 1203 
\bibitem{} Thorne, K.\ S., Zytkow, A.\ N.\ 1975, ApJ 199, L19
\bibitem{} Thorne, K.\ S., Zytkow, A.\ N.\ 1977, ApJ 212, 832
\bibitem{} Timmes, F.~X., Woosley, S.~E., Weaver, T.~A.\ 1996, ApJ
	   457, 834 
\bibitem{} Tutukov, A.\ V., Chugaj, N.\ N.,  Yungelson, L.\ R. 1984, PAZh 10, 586
\bibitem{} Tutukov, A.\ V., Yungelson, L.\ R, 1993, AZh 70, 812
\bibitem{} van~den Heuvel, E.~P.~J., Habets, G.\ 1984, Nat 309, 598
\bibitem{} van~den Heuvel, E.~P.~J., Rappaport, S.\ 1987, in Physics
of Be stars, Cambride Univ. Press, p.\ 291 
\bibitem{} van den Heuvel, E.~P.J., van Paradijs J.\ 1997, ApJ 483, 399
\bibitem{} van Kerkwijk, M.\, H., Kulkarni, S.\, R.\, 1999, ApJ {\em in press}
\bibitem{} Woosley, S.~E.~1987, The origin and evolution of neutron
stars; Proceedings of the IAU Symposium, Dordrecht, D. Reidel
Publishing Co., p.\ 255 
\bibitem{} Woosley, S.~E.~ Weaver, T.\ A.\ 1992, in The structure and evolution
neutron stars, eds.\ D.~Pines, R.~Ramagaki and S.~Tsuruta, Addison Wesley,
Redwood City, PA, p.~235 
\end{thebibliography}
\end{document}